\documentclass{article}
\usepackage{spconf,amsmath,graphicx}
\usepackage{amssymb}
\usepackage{url}
\usepackage{cite}
\usepackage{booktabs}
\usepackage{color}
\usepackage{microtype}
\usepackage{multirow}

\DeclareMathOperator*{\argmin}{argmin}
\DeclareMathOperator*{\median}{median}

\newcommand\startWithCleanPageOptionally{}

\title{Interference Reduction in Music Recordings Combining Kernel Additive Modelling and Non-Negative Matrix Factorization}
\name{Delia Fano Yela$^1$, Sebastian Ewert$^1$, Derry FitzGerald$^2$, Mark Sandler$^1$\thanks{This work was funded by EPSRC grant EP/L019981/1.}}
\address{Queen Mary University of London, UK$^1$\\Nimbus Centre, Cork Institute of Technology, Ireland$^2$}
\begin{document}
\ninept

\linespread{0.94}
\selectfont
\maketitle
\begin{abstract}

In live and studio recordings unexpected sound events often lead to interferences in the signal. For non-stationary interferences, sound source separation techniques can be used to reduce the interference level in the recording. In this context, we present a novel approach combining the strengths of two algorithmic families: NMF and KAM. The recent KAM approach applies robust statistics on frames selected by a source-specific kernel to perform source separation. Based on semi-supervised NMF, we extend this approach in two ways. First, we locate the interference in the recording based on detected NMF activity. Second, we improve the kernel-based frame selection by incorporating an NMF-based estimate of the clean music signal. Further, we introduce a temporal context in the kernel, taking some musical structure into account. Our experiments show improved separation quality for our proposed method over a state-of-the-art approach for interference reduction.

 \end{abstract}
\begin{keywords}
Source separation, Kernel Additive Modelling, Non-Negative Matrix Factorization, Interference Reduction.
\end{keywords}
\section{Introduction}
\label{sec:intro}

In professional music recordings one often has to deal with various types of sound interferences. For example, a person in the audience experiencing a coughing fit during a classical music concert can be a major disturbance. Similarly, fans screaming too close to one of the stage microphones can render the entire channel useless in post-production. Further, studio sessions are often subject to strict time budgets and thus many tracks are only recorded until the sound engineer assesses the last take to be good enough -- only to find a door being slammed or an object falling on the floor in this one good take during the actual production.

The difficulty of removing such an interference strongly depends on its type. Stationary interferences, such as mains or fluorescent light hum, can often already be reduced by simple (Wiener) filtering techniques \cite{Hayes96_StatisticalDSP_Wiley}. Non-stationary interferences such as the ones described above, however, require more complex signal models and sound source separation techniques to differentiate noise from non-noise signal components. In this context, Non-Negative Matrix Factorization (NMF) proves to be a powerful tool and most state-of-the-art source separation methods are based on NMF variants \cite{CichockiZP09_AlternateAlgorithmsNmf_Book}. The basic idea behind NMF is to model a time-frequency representation of the signal as a product of two matrices. The columns of the first matrix are often interpreted as \emph{templates} capturing the spectral properties of the individual sound sources in the signal; the rows of the second matrix are often referred to as the corresponding \emph{activations}, encoding when and how strong each template is active in the input signal.

Applying the original NMF approach \cite{LeeS00_AlgorithmsNmf_NIPS} to audio and music data, however, was found to rarely yield useful results \cite{FitzGeraldCC08_ExtendedNMFVariants_CIN}. Therefore, various extensions were proposed integrating various constraints on the parameter estimation process. Examples include sparsity and temporal continuity constraints \cite{Virtanen07_MonauralSoundSourceSeparation_TASLP} or harmonicity constraints \cite{BertinBV10_EnforcingHarmonicityInBayesNMF_TASLP}. Further, various types of side information have been used, such as user-assisted annotations \cite{Smaragdis09_UserGuidedAudioSelection_ACM-UIST} and musical score information \cite{EwertPMP14_ScoreInformedSourceSep_IEEE-SPM}. One of the most widely used and successful approaches is to employ training data (\emph{Supervised NMF}): using recordings containing only a single sound source, corresponding templates representing that source can easily be computed \cite{ChoC04_LearningNMFBases_Interspeech}. This way, one can avoid relying on specific assumptions about the statistical independence of the sources \cite{AbdallahP04_TranscriptionViaNMF_ISMIR}. 
As a major drawback of this approach, however, the quality of the separation result heavily depends on the assumption that the acoustical conditions in the training material and in the recording to be processed are similar. The more this assumption is violated, the more artefacts are to be expected.

As an alternative to NMF, \emph{Kernel Additive Modelling} (KAM) \cite{LiutkusFRPD14_KernelAdditive_IEEE-TSP} was proposed for various tasks in source separation, e.g.\,singing voice separation \cite{FitzGerald12_MedianVocal_ISSC,RafiiPardo13_REPET_IEEE-TASLP}, the separation of harmonic from percussive signal components \cite{Fitzgerald10_HarmPercSep_DAFX} or the reduction of microphone bleeding in multi-channel recordings \cite{PraetzlichBLM15_InterferenceReduction_ICASSP}.
In general, the idea behind KAM is to exploit that the magnitude of a bin in a time-frequency representation is often similar or related to the magnitude of certain other bins -- which bins are similar is described by a so called kernel. If the magnitude of a given bin deviates in an unexpected way from the bins defined in the kernel, one can assume that this bin is overlaid by another sound source and we can use the kernel bins to reconstruct the overlaid one. Since some of the kernel bins might be overlaid by other sounds as well, or are not exact repetitions, one uses \emph{Robust Statistics}, in particular order statistics, to identify the commonalities between the bins while neglecting the outliers. 

To apply a KAM-based method to a source separation problem, one needs to design a corresponding kernel that identifies similar spectral bins for the sources we want to keep while ignoring the energy associated with other sources. In existing KAM approaches, this kernel design is often rather rudimentary. For example, to eliminate the singing voice from recordings, the methods proposed in \cite{FitzGerald12_MedianVocal_ISSC,RafiiPardo13_REPET_IEEE-TASLP} assume that the accompaniment playing the harmony changes more slowly than the singing voice and thus that there are many frames with similar accompaniment. The kernel used in \cite{FitzGerald12_MedianVocal_ISSC,RafiiPardo13_REPET_IEEE-TASLP} is simply a function finding the $K$ most similar frames based on the Euclidean distance. However, using such simple kernels one implicitly assumes that the energy in frames will be dominated by the sound source we want to keep -- otherwise the similarity measure fails to identify similar frames. Therefore, while standard KAM is free of the need for suitable training data as in supervised NMF, it might fail to find similar frames if the signal-to-interference ratio is low. In particular, with sudden, loud interferences as to be expected in our application scenario, existing KAM approaches are likely to fail.

Our main idea is to combine the strengths of both approaches. In particular, while training data in supervised NMF might not be precise enough to yield a high quality signal model as needed for source separation, it might be discriminative enough to obtain an initial signal model for the music, which can be used to design an adaptive, interference-resilient kernel for KAM. More precisely, we let the user provide keywords to describe the interference (e.g.`cough') and retrieve corresponding training data from the publicly available freesound\footnote{\url{https://www.freesound.org/}} archive. After computing templates specific for the interference from the training data, we apply a semi-supervised NMF, i.e. we fix templates for the interference and learn some additional free templates to model the music from the actual input signal. Then, using the (HMM-smoothed) NMF activations for the fixed interference templates, we automatically locate the interference within the recording -- this way, in contrast to existing KAM approaches, we can filter the signal only where needed. Second, using the activations for the free templates, we can reconstruct an initial rough estimate for the music, where the interference is strongly reduced as most of the corresponding energy is already captured by the interference templates. Based on this initial model, we identify for each frame affected by the interference a list of similar frames, which are then used within the KAM framework to produce the final output. As additional contributions, we modify the standard kernels used in KAM by incorporating a temporal context into the similarity search which essentially yields a simple regularizer promoting temporal continuity of the kernels across frames, as well as a smoothing technique, which enhances the method's invariance against small variations in the fundamental frequency.

The remainder of the paper is organized as follows. In Section~\ref{sec:proposedMethod} we describe the technical details of our approach. Next, in Section~\ref{sec:experiments} we compare our proposed method with standard KAM and semi-supervised NMF in a series of systematic experiments. Finally, we conclude the paper in Section~\ref{sec:conclusion} with an outlook on future work.

\startWithCleanPageOptionally
\section{Proposed Method}
\label{sec:proposedMethod}

Overall, we develop our method as an extension to \emph{Kernel Additive Modelling (KAM)} \cite{LiutkusFRPD14_KernelAdditive_IEEE-TSP}. From a modelling point of view, KAM and the more widely known Gaussian Processes (GP) share similar concepts. In both cases, the idea is that for many signals we can estimate the value of a single sample by looking at the value of neighbouring samples. For example, a low frequency signal corrupted by white noise can be reconstructed by averaging the values of neighbouring samples. This operation is essentially similar to a low-pass FIR filter, just that KAM and GP enable the use of much more general notions of similarity or neighbourhood. KAM differs from GP in several aspects. First, the similarity kernel in KAM can depend on the observations themselves \cite{QuinoneroCandelaR2005_UnifyingViewGP_JMLR}, which we exploit in the following. Second, non-Gaussian noise corrupting the sample values can be modelled. Third, as an instance of kernel local regression, KAM does not require the inversion of a data covariance matrix (as in GPs), which typically leads to considerable improvement in terms of computational costs \cite{LiutkusFRPD14_KernelAdditive_IEEE-TSP}.

\begin{figure}[t]
\centering
\includegraphics[width=\columnwidth,height=8cm]{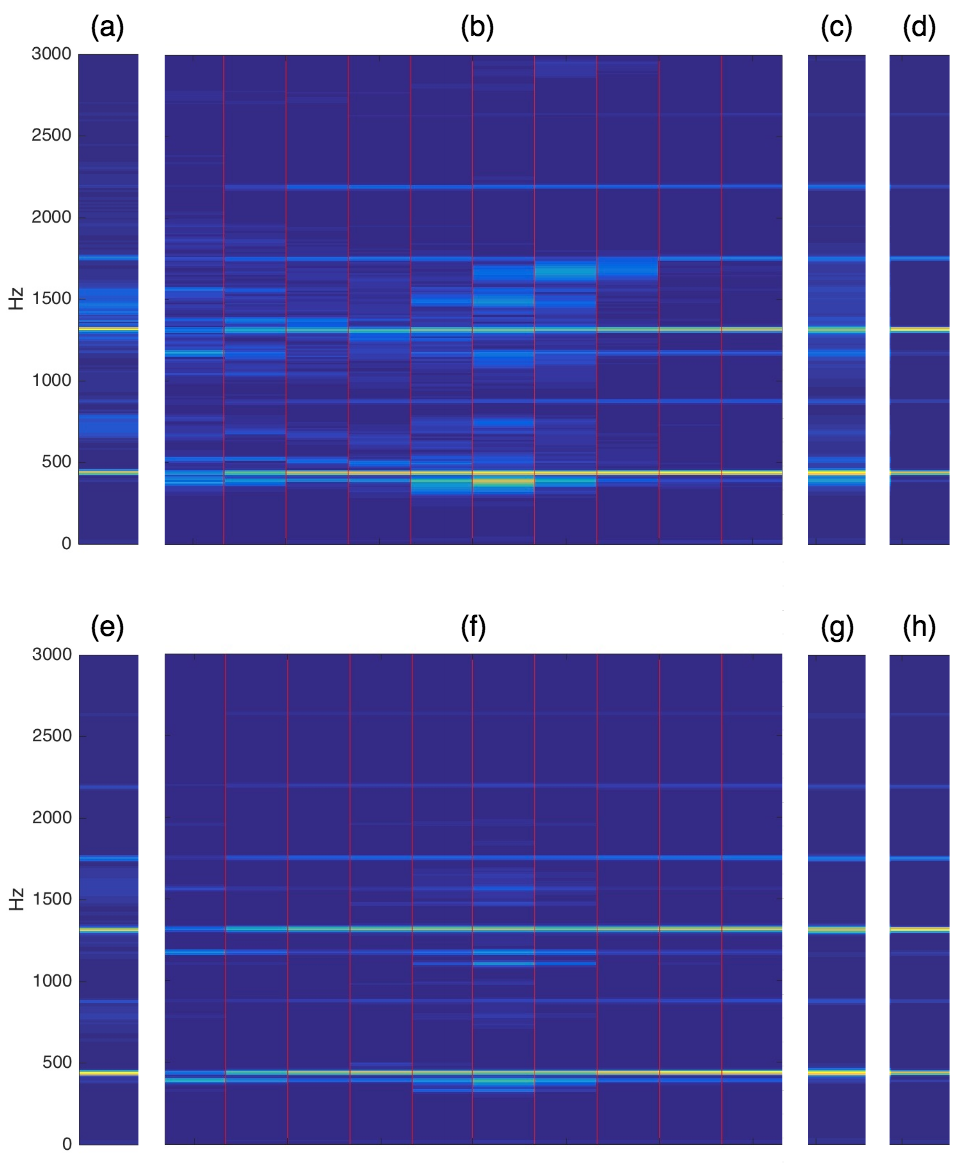}
\vspace{-0.5cm}
\caption{Individual steps in our proposed method, (a)-(d) using standard KAM, (e)-(h) using our proposed extension: (a,e) current frame used for similarity search, (b,f) first 10 closest frames found, (c,g) estimated frame and (d,h) ideal clean frame.}
\label{fig:methodIllustration}
\vspace{-0.3cm}
\end{figure}

The KAM framework as a whole is relatively rich, both in possible application scenarios and theory. Due to space constraints, we will only present a smaller subset that was also used in a similar form in the REPET family of methods for singing voice removal \cite{RafiiPardo13_REPET_IEEE-TASLP}. To this end let $x$ be the signal to be processed with $x(t) = s(t) + n(t)$, where $s$ and $n$ are the clean music and the interference signal, respectively. Further, let
$X, S \in \mathbb{C}^{F \times T}$ be the spectrograms of $x$ and $s$. 
In the following, we exploit that spectral frames in $S$ typically occur several times in similar form, either because note constellations are repeated over time (as is common in music) or because notes are being held for a while. The interference on the other hand may or may not be repetitive and thus we cannot make any assumptions here. Therefore, we will model only $s$ in KAM without considering the interference $n$ as an actual sound source but just as noise with an unknown distribution. Since $s$ only consists of a single channel, we can eliminate many unnecessary elements in KAM (multi-channel and iterative re-estimation extensions, compare \cite{LiutkusFRPD14_KernelAdditive_IEEE-TSP}), resulting in a very simple representation. More precisely, let $\mathcal{I}: F\times T \rightarrow \mathcal{P}(F\times T)$ be a similarity kernel function that assigns to every time-frequency bin $(f,t)$ a list of $K$ similar bins, i.e.\,$\forall (f,t) \in F\times T: |\mathcal{I}(t,f)| = K$. As in the case of REPET, we use a frame-wise, $K$-nearest neighbours ($K$-NN) function based on the Euclidean distance, i.e. $(f,\tilde{t})$ is in $\mathcal{I}(f,t)$ if frame $\tilde{t}$ is among the $K$ most similar frames. This process is illustrated in Fig.~\ref{fig:methodIllustration}, where for a given frame shown in Fig.~\ref{fig:methodIllustration}a the $K=10$ most similar frames are shown in Fig.~\ref{fig:methodIllustration}b.

Once this notion of similarity between bins is established, we can try to calculate a noise-free estimate for each bin $(f,t)$ from the bins in $\mathcal{I}(f,t)$. In KAM  \cite{LiutkusFRPD14_KernelAdditive_IEEE-TSP} this goal is expressed as an optimization problem over so called \emph{model cost functions} $\mathcal{L}$. More precisely, we get\footnote{Note that the formal requirement to have noisy data extracts for $S$ (as used in KAM) are directly given here as $X(f,\tilde{t})$. This is the result of having only a single sound source in a single channel, which makes the iterative re-estimation in KAM unnecessary and eliminates all elements related to the estimation of the mixing matrix, compare \cite{LiutkusFRPD14_KernelAdditive_IEEE-TSP}.}:
\[
\overline{S}(f,t) = \argmin_{\lambda \in \mathbb{R}} \sum_{ (f,\tilde{t}) \in \mathcal{I}(f,t)}{\mathcal{L}(\overline{X}(f,\tilde{t}),\lambda)},
\]
where $\overline{X}$ is the magnitude of $X$. Here $\mathcal{L}$ models our belief regarding how good or bad a specific choice for $\overline{S}(f,t)$ is, considering that we call all elements in $\mathcal{I}(f,t)$ similar to it. A common choice in the KAM framework is $\mathcal{L}(a,b) := |a-b|$. This choice is interesting for two reasons. First, it expresses from a probabilistic point of view that we expect some larger deviations in the difference $a$ and $b$, and that this distance is not Gaussian distributed (otherwise a Euclidean distance would be optimal here). Second, this choice leads us to the use of robust statistics in the form of the median, which as an operator is invariant against outliers (breakdown point is $50$\%) and thus allows robust parameter estimation in the presence of noise. More precisely, with $\mathcal{L}(a,b) := |a-b|$ the solution to the above problem is:
\[
\overline{S}(f,t) = \median( \overline{X}(f,\tilde{t})  | (f,\tilde{t}) \in \mathcal{I}(f,t)).
\]
The result is shown in Fig.~\ref{fig:methodIllustration}c.

Comparing the results of this approach shown in Fig.~\ref{fig:methodIllustration}c with the clean signal in Fig.~\ref{fig:methodIllustration}d, we can observe an example of when this approach fails. In particular, comparing Fig.~\ref{fig:methodIllustration}a and Fig.~\ref{fig:methodIllustration}d, we see that the input frame is overlaid by a strong interference. With such a low signal-to-noise ratio (SNR), the interference dominates in the similarity search based on the Euclidean distance and the kernel function $\mathcal{I}(f,t)$ points to too many noisy examples, which even the median operation cannot eliminate. In particular, despite being strongly invariant against outliers from a robust statistics point of view, the outliers cannot be identified anymore based on a selection of frames as shown in Fig.~\ref{fig:methodIllustration}c.

Our idea is now to improve the $K$-NN search in KAM in several ways, making the kernel function more invariant against the interference signal.
To this end, we build a first initial signal model based on NMF using training data. While the training data might differ from the actual interference signal, and thus an actual source separation based on this method would yield results of low quality, it might be good enough to gather more information about the signal and reduce the influence of the interference. More precisely, similar to \cite{ElBadawyDO2014_FreeSoundSourceSerp_MLSP}
we let the user provide keywords to describe the interference (e.g.`cough') and retrieve corresponding example recordings from the freesound archive.
Concatenating these recordings into a single file, we compute its magnitude spectrogram $\overline{X}_N$ as well as an NMF factorization $\overline{X}_N \approx W_N H$
using the well-known Lee-Seung NMF updates for the generalized Kullback-Leibler divergence $D_{\text{KL}}$ \cite{LeeS00_AlgorithmsNmf_NIPS}, i.e. we minimize $D_{\text{KL}}(\overline{X}_N, W_N H)$ over non-negative matrices $W_N$ and $H$. The only parameter here is the NMF rank $R_1$. After this, the columns of $W_N$ contains templates reflecting the spectral properties of the interference signal. 

In a next step, we employ NMF to model our input spectrogram $\overline{X}$ using a combination of interference templates, $W_N$, and music templates, $W_S$.
Here, the interference templates can be kept fixed and we only need to learn the music templates, which is often referred to as \emph{semi-supervised NMF}.
More precisely, we minimize the function $D_{\text{KL}}(\overline{X}, W_N H_N + W_S H_S)$ over $H_N$, $W_S$ and $H_S$ (i.e. we fix $W_N$). In this case, the update rules are similar to regular NMF:
\[
H_N \leftarrow H_N \odot \frac{W_N^\top \mathcal{R}}{W_N^\top \cdot J}
\quad \text{and} \quad
H_S \leftarrow H_S \odot \frac{W_S^\top \mathcal{R}}{W_S^\top \cdot J},
\]
\[
W_S \leftarrow W_S \odot \frac{\mathcal{R} H_S^\top}{J \cdot H_S^\top},
\quad \text{with} \quad
\mathcal{R} := \frac{\overline{X}}{ W_N H_N + W_S H_S}
\]
and $J$ the all-one matrix. After convergence, the rows of $H_N$ capture the activations of the interference templates, while $W_S H_S$ yields an approximation of the magnitude spectrogram for the music. Using these two interpretations, we employ these results for two different purposes. First, we use $H_N$ to identify where the interference is, which will enable us to filter only frames with interference (in contrast to regular KAM). To this end, we sum the values in $H_N$ in each frame to obtain a single curve indicating interference activity, which we decode using an HMM, resulting in a binary, frame-wise interference indicator vector $I$. The parameters of the HMM implemented, detection threshold and cost of changing state, were adjusted to favour recall over precision in the detection.

Next, we exploit that while the interference templates might not perfectly reflect the properties of the target interference (and thus a separation based on this model would be of low quality), they do capture typically a considerable amount of interference energy in the signal. Therefore, we can improve the $K$-NN search in KAM kernel by replacing the input spectrogram $\overline{X}$ containing the interference with the NMF approximation for the music $\widetilde{X} := W_S H_S$. The resulting improvement is clearly visible in Fig.~\ref{fig:methodIllustration}. Replacing the $\overline{X}$-frame (Fig.~\ref{fig:methodIllustration}a) with the corresponding $\widetilde{X}$-frame (Fig.~\ref{fig:methodIllustration}e) in the similarity search, we see that the frames selected as nearest neighbours (Fig.~\ref{fig:methodIllustration}f) are much closer to the actual target (Fig.~\ref{fig:methodIllustration}d = Fig.~\ref{fig:methodIllustration}h). The median filter can then remove remaining noise robustly, bringing the result (Fig.~\ref{fig:methodIllustration}g) much closer to the target (Fig.~\ref{fig:methodIllustration}h). 

However, in particular if musical patterns are rarely or not repeated in the mixture, we observed that sometimes the frames selected as nearest neighbours using $\widetilde{X}$ still contained a significant amount of interference energy, again potentially rendering the median filtering ineffective. As a further extension, we therefore propose to check if a frame selected as nearest neighbour was previously already identified as an interference frame and, in that case, use the corresponding $\widetilde{X}$-frame for the median filtering instead of the $\overline{X}$-frame. As it is shown in Section~\ref{sec:experiments}, this extension additionally reduces the interference impact on the separation result.

A further problem we observed is that the kernel $\mathcal{I}$ was often changing considerably between frames in the sense that often $(f,\tilde{t}) \in \mathcal{I}(f,t)$ would not imply $(f,\tilde{t}+1) \in \mathcal{I}(f,t+1)$. Without this property, however, we observed a slight pitch jitter in the magnitude across frames after median filtering, which was audible in the final time domain signal.
To further temporally stabilize the kernel function, we propose incorporating a temporal context into the similarity search. More precisely, instead of comparing frames $t$ and $\tilde{t}$ with a simple squared Euclidean distance $\sum_f{ (\widetilde{X}(f,t) - \widetilde{X}(f,\tilde{t}))^2}$, we employ
\vspace{-0.35cm}
\[
\sum_f{ \sum_{c=-C}^C (\widetilde{X}(f,t+c) - \widetilde{X}(f,\tilde{t}+c))^2}
\vspace{-0.1cm}
\]
as frame distance in the $K$-NN search, where $C$ specifies the \emph{temporal extent}. We found this simple extension to act as a surprisingly effective temporal regularizer for $\mathcal{I}$. Further, we found that filtering $\widetilde{X}$ slightly in frequency direction before the $K$-NN search using a small Gaussian kernel additionally improved the results, as it makes the similarity search invariant to small changes in the fundamental frequency of harmonic sounds.

To perform the actual separation, we employ soft masking (similar to Wiener filtering). In particular, our method yields an estimate $\overline{S}$ for the magnitude spectrogram of the music. We define a corresponding estimate for the noise, here interference, as $\overline{N} = \max(\overline{X} - \overline{S},0)$. This way, we can obtain an estimate $S$ for the complex music spectrogram via
$S = \frac{\overline{S}}{\overline{N} + \overline{S}} \odot X$. Overall, we found our method combining NMF and KAM to improve over both approaches considerably, which we demonstrate in the next section.

\startWithCleanPageOptionally
\section{Experiments}
\label{sec:experiments}
We evaluated our proposed method using freely available recordings, in particular interferences and instrumental solo stems from multitrack recordings \cite{BittnerSTMCB_MedleyDB_ISMIR}. We chose interferences that typically occur in a live or studio scenario including cough sounds, door slams, sounds of objects of different material being dropped, chair-drag sounds as well as audience screams. The music dataset contains 58 instrumental mono stems from the multitrack MedleyDB dataset \cite{BittnerSTMCB_MedleyDB_ISMIR}, covering 23 different instruments ranging from guitar, violin, piano over to bass, trombone or flute.

Similar to \cite{ElBadawyDO2014_FreeSoundSourceSerp_MLSP}, we retrieved recordings of interferences from freesound.org -- this way, the method does not rely on the availability of non-public training data and is easily extended to other types of interferences. However, this also implies that the quality and number of training samples can vary, and thus explains why, in our case, each interference has a different amount of training data, ranging from 10 scream samples to 40 coughs tracks. The separation quality is expected to improve as the number of tracks in the training data increases.  

We created test recordings by making artificial linear mixes of stems and test interference recordings independent of the training data and of each other (other acoustic conditions). In order to achieve a controlled mix of instrumental and interference levels, all tracks were normalised to a specific RMS energy. Then three interferences are added to the music at different SNR, measured on the segment where the interference is active. The final mix is a 30s long monaural recording with three different sounds of the same kind interfering at different times at a certain SNR. 

We evaluated the proposed method on the resulting 290 mixtures (58 instrumental stems times 5 types of interferences), measuring the separation performance using the BSS Eval toolbox \cite{VincentGF06_PerformanceMeasurement_IEEE-TASLP}, obtaining a Signal-to-Distortion Ratio (SDR) and Signal-to-Interference Ratio (SIR) for each mixture separation. The SDR is used as a measure to indicate the overall separation performance, whereas the SIR shows how much of the interference signal is left in the signal estimate. To indicate the improvement over the raw music-interference mix, we employ the normalized SDR/SIR (NSDR/NSIR) as in \cite{LiutkusDR15_KAML_ICASSP}, i.e. from the SDR obtained using our method we subtract the SDR from the mix. This way, we can account for the fact that a separation at a low SNR is more difficult than at a high SNR, making results for different SNRs more comparable. 

Here we have chosen supervised-NMF to represent the current state-of-the-art method to quantitatively compare its separation performance to the proposed method. In order to obtain a competitive baseline, we use the same learned dictionary for both methods and we also optimise the NMF rank  with a parameter sweep. Tables~\ref{tab:resultsKamVsNmf} and \ref{tab:resultsIndividualExtensions} show the overall results, averaged across all NSDR/ NSIR values of every mixture, for our proposed method as well as for the semi-supervised NMF approach. 
Comparing the results, our proposed method yields a higher separation quality than the NMF-based method not only for a 0dB SNR mixture, but also for mixtures where the interference is 3dB and 6dB below the instrumental RMS energy. Overall, we obtain an improvement between $1.4$ and $2.0$dB, which from a relative point of view is quite considerable.

\begin{table}\centering
{\footnotesize
\begin{tabular}{@{}rrrrcrrrcrrr@{}}\toprule
& \multicolumn{3}{c}{NSDR} & \phantom{abc}& \multicolumn{3}{c}{NSIR} \\
\cmidrule{2-4} \cmidrule{6-8} 
& $0$dB & $-3$dB & $-6$dB && $0$dB & $-3$dB & $-6$dB \\ \midrule
Prop. & 6.78 & 4.76 & 2.52 && 16.79 & 15.30 & 13.69 \\
NMF &  4.76 & 3.16 & 1.13 && 13.15 & 14.40 & 15.62\\
\bottomrule
\end{tabular}}
\caption{Comparison of our method with supervised NMF for different SNR values.}
\label{tab:resultsKamVsNmf}
\end{table}

In order to measure the influence of the individual components of our proposed method, Table~\ref{tab:resultsKamVsNmf} shows results separately for several variations of our method. To provide another angle on the results and focus on the positions where the interferences actually happen, we evaluated the separation performance by averaging across the three segments in the mix where the interference is active, and so the resulting NSDR scores are not directly comparable to Table~\ref{tab:resultsKamVsNmf}.

Starting with a baseline KAM approach as described in \cite{FitzGerald12_MedianVocal_ISSC}, \textit{Variant V1} adds the NMF interference detection step introduced in Section~\ref{sec:proposedMethod}. The high NSDR shows the interference was successfully identified and reduced. 
\textit{Variant V2} further adds the improved similarity measure of our proposed method, where similarity is measured based on a rough NMF estimate of the signal. Additionally, the frame-wise similarity search used in standard KAM (and \textit{Variant~V1}) is modified to account for the local temporal context in V2 as introduced in Section~\ref{sec:proposedMethod}. The higher NSDR shows that the temporal context stabilizes not only the kernel but also the results. In this context, it is important to remark that our test signal are only 30 seconds long -- for longer signals with additional repetitions of musical patterns, we would expect even higher improvements in NSDR.
Overall, both extensions improve the capability of our method to better identify and select similar frames and thus to increase the performance of the median filtering step.

\begin{table}\centering
{\footnotesize
\begin{tabular}{lrr} \toprule
& NSDR & NSIR \\ \midrule
V1: Standard KAM + & \multirow{2}{*}{7.09} & \multirow{2}{*}{13.62}\\
\hspace{0.5cm}NMF Interference Detection & &\\
V2: V1 + NMF-based Kernel Similarity + & \multirow{2}{*}{7.92} & \multirow{2}{*}{15.48}\\
\hspace{0.5cm}Temporal Context& &\\
V3: V2 + Adaptive Frame Selection + & \multirow{2}{*}{8.84} & \multirow{2}{*}{14.53}\\
\hspace{0.5cm}Smoothing (Proposed Method) & &\\
\bottomrule
\end{tabular}}
\caption{Influence of individual KAM extensions on the separation result (interference at 0dB SNR; separation evaluated on the segments affected by an interference).}
\label{tab:resultsIndividualExtensions}
\vspace{-0.5cm}
\end{table}

\textit{Variant V3} is an extension of \textit{Variant V2} incorporating the smoothing filter and the adaptive frame selection, which replaces frames in the median filter in which an inferences was detected with the corresponding frames from the NMF estimate, see Section~\ref{sec:proposedMethod}.
As shown in Table~\ref{tab:resultsIndividualExtensions}, both extensions further improve the NSDR over variant \textit{Variant V2}. However, the NSIR values are sometimes lower -- in our experiments, we found this to be a side effect of the smoothing filter, which slightly blurs the spectrum, leading to a tendency of leaving more residual energy in the output.
However, overall, these results show that each of our proposed extensions measurably improves the separation quality.

\startWithCleanPageOptionally
\section{Conclusion}
\label{sec:conclusion}
We have presented a new method for interference reduction combining NMF and KAM. Our method exploits advantages of both techniques: using a spectral dictionary we detect the interference occurrences and produce an initial clean signal estimate using NMF. This estimate is used to improve the similarity measure used in KAM, making it less dependent on the SNR of the interference. A further extension incorporates a temporal context into the similarity search, which stabilized the KAM kernel function and further improved the separation results. Finally, an adaptive frame selection mechanism replacing frames with interferences with corresponding NMF-estimates in the median filter led to an additional improvement, in particular for short recordings. For solo instrumental recordings, our experiments showed a considerable improvement in separation quality for our proposed method over a competitive method based on supervised NMF. Possible future directions for extending this work would include an improved similarity search as well as the implementation of source-specific kernels both in time and frequency direction.

\startWithCleanPageOptionally
\bibliographystyle{IEEEbib}
\bibliography{referencesMusic}

\begin{thebibliography}{10}

\bibitem{Hayes96_StatisticalDSP_Wiley}
Monson~H. Hayes,
\newblock {\em Statistical Digital Signal Processing and Modeling},
\newblock Wiley, 1st edition, 1996.

\bibitem{CichockiZP09_AlternateAlgorithmsNmf_Book}
Andrzej Cichocki, Rafal Zdunek, and Anh~Huy Phan,
\newblock {\em Nonnegative Matrix and Tensor Factorizations: Applications to
  Exploratory Multi-Way Data Analysis and Blind Source Separation},
\newblock John Wiley and Sons, 2009.

\bibitem{LeeS00_AlgorithmsNmf_NIPS}
Daniel~D. Lee and H.~Sebastian Seung,
\newblock ``Algorithms for non-negative matrix factorization,''
\newblock in {\em Advances in Neural Information Processing Systems}, Denver,
  CO, USA, 2000, pp. 556--562.

\bibitem{FitzGeraldCC08_ExtendedNMFVariants_CIN}
Derry FitzGerald, Matt Cranitch, and Eugene Coyle,
\newblock ``Extended nonnegative tensor factorisation models for musical sound
  source separation (article id 872425),''
\newblock {\em Computational Intelligence and Neuroscience}, vol. 2008, 2008.

\bibitem{Virtanen07_MonauralSoundSourceSeparation_TASLP}
Tuomas Virtanen,
\newblock ``Monaural sound source separation by nonnegative matrix
  factorization with temporal continuity and sparseness criteria,''
\newblock {\em IEEE Transactions on Audio, Speech and Language Processing},
  vol. 15, no. 3, pp. 1066--1074, 2007.

\bibitem{BertinBV10_EnforcingHarmonicityInBayesNMF_TASLP}
Nancy Bertin, Roland Badeau, and Emmanuel Vincent,
\newblock ``Enforcing harmonicity and smoothness in {B}ayesian non-negative
  matrix factorization applied to polyphonic music transcription,''
\newblock {\em IEEE Transactions on Audio, Speech, and Language Processing},
  vol. 18, no. 3, pp. 538--549, 2010.

\bibitem{Smaragdis09_UserGuidedAudioSelection_ACM-UIST}
Paris Smaragdis,
\newblock ``User guided audio selection from complex sound mixtures,''
\newblock in {\em Proceedings of the {ACM} Symposium on User Interface Software
  and Technology ({UIST})}, New York, NY, USA, 2009, pp. 89--92.

\bibitem{EwertPMP14_ScoreInformedSourceSep_IEEE-SPM}
Sebastian Ewert, Bryan Pardo, Meinard M{\"u}ller, and Mark~D. Plumbley,
\newblock ``Score-informed source separation for musical audio recordings: An
  overview,''
\newblock {\em IEEE Signal Processing Magazine}, vol. 31, no. 3, pp. 116--124,
  May 2014.

\bibitem{ChoC04_LearningNMFBases_Interspeech}
Yong-Choon Cho and Seungjin Choi,
\newblock ``Learning nonnegative features of spectro-temporal sounds for
  classification,''
\newblock in {\em In Proceedings of InterSpeech}, 2004.

\bibitem{AbdallahP04_TranscriptionViaNMF_ISMIR}
Samer Abdallah and Mark Plumbley,
\newblock ``Polyphonic transcription by non-negative sparse coding of power
  spectra,''
\newblock in {\em Proceedings of the International Society for Music
  Information Retrieval Conference (ISMIR)}, Barcelona, Spain, 2004, pp.
  318--325.

\bibitem{LiutkusFRPD14_KernelAdditive_IEEE-TSP}
Antoine Liutkus, Derry FitzGerald, Zafar Rafii, Bryan Pardo, and Laurent
  Daudet,
\newblock ``Kernel additive models for source separation,''
\newblock {\em {IEEE} Transactions on Signal Processing}, vol. 62, no. 16, pp.
  4298--4310, 2014.

\bibitem{FitzGerald12_MedianVocal_ISSC}
Derry FitzGerald,
\newblock ``Vocal separation using nearest neighbours and median filtering,''
\newblock in {\em Proceedings of the Irish Signals and Systems Conference
  (ISSC)}, 2012, pp. 1--5.

\bibitem{RafiiPardo13_REPET_IEEE-TASLP}
Zafar Rafii and Bryan Pardo,
\newblock ``Repeating pattern extraction technique ({REPET}): A simple method
  for music/voice separation.,''
\newblock {\em {IEEE} Transactions on Audio, Speech, and Language Processing},
  vol. 21, no. 1, pp. 71--82, 2013.

\bibitem{Fitzgerald10_HarmPercSep_DAFX}
Derry FitzGerald,
\newblock ``Harmonic/percussive separation using median filtering,''
\newblock in {\em Proceedings of the International Conference on Digital Audio
  Effects ({DAFx})}, Graz, Austria, 2010, pp. 246--253.

\bibitem{PraetzlichBLM15_InterferenceReduction_ICASSP}
Thomas Pr{\"a}tzlich, Rachel Bittner, Antoine Liutkus, and Meinard M{\"u}ller,
\newblock ``Kernel additive modeling for interference reduction in
  multi-channel music recordings,''
\newblock in {\em Proceedings of the {IEEE} International Conference on
  Acoustics, Speech, and Signal Processing ({ICASSP})}, Brisbane, Australia,
  2015, pp. 584--588.

\bibitem{QuinoneroCandelaR2005_UnifyingViewGP_JMLR}
Joaquin Qui{\~n}onero-Candela and Carl~Edward Rasmussen,
\newblock ``A unifying view of sparse approximate gaussian process
  regression,''
\newblock {\em Journal of Machine Learning Research}, vol. 6, no. Dec, pp.
  1939--1959, 2005.

\bibitem{ElBadawyDO2014_FreeSoundSourceSerp_MLSP}
Dalia El~Badawy, Ngoc~QK Duong, and Alexey Ozerov,
\newblock ``On-the-fly audio source separation,''
\newblock in {\em Proceedings of the IEEE International Workshop on Machine
  Learning for Signal Processing (MLSP)}, 2014, pp. 1--6.

\bibitem{BittnerSTMCB_MedleyDB_ISMIR}
Rachel~M. Bittner, Justin Salamon, Mike Tierney, Matthias Mauch, Chris Cannam,
  and Juan~Pablo Bello,
\newblock ``Medleydb: {A} multitrack dataset for annotation-intensive {MIR}
  research,''
\newblock in {\em Proceedings of the International Society for Music
  Information Retrieval Conference ({ISMIR})}, Taipei, Taiwan, October 2014,
  pp. 155--160.

\bibitem{VincentGF06_PerformanceMeasurement_IEEE-TASLP}
Emmanuel Vincent, R\'{e}mi Gribonval, and C\'{e}dric F\'{e}votte,
\newblock ``Performance measurement in blind audio source separation,''
\newblock {\em IEEE Transactions on Audio, Speech, and Language Processing},
  vol. 14, no. 4, pp. 1462--1469, 2006.

\bibitem{LiutkusDR15_KAML_ICASSP}
Antoine Liutkus, Derry Fitzgerald, and Zafar Rafii,
\newblock ``Scalable audio separation with light kernel additive modelling,''
\newblock in {\em IEEE International Conference on Acoustics, Speech and Signal
  Processing (ICASSP)}, 2015, pp. 76--80.

\end{thebibliography}

\end{document}